\documentstyle[multicol,aps,prl,twocolumn]{revtex}
\newif\iffigs
\figstrue
\iffigs
\input{psfig}
\fi
\begin{document}
\twocolumn[\hsize\textwidth\columnwidth\hsize\csname@twocolumnfalse\endcsname

\title{Dispersive stabilization of the inverse cascade for the
Kolmogorov flow} \author{B. Legras,$^1$\cite{email} B. Villone,$^2$
and U. Frisch$^3$} \address{$^1$ CNRS, LMD/ENS, 24 rue Lhomond, 75231
Paris Cedex 5, France\\ $^2$ Istituto di Cosmogeofisica, CNR, C. Fiume
4, 10133 Torino, Italy\\ $^3$ CNRS UMR 6529, Observatoire de la C\^ote
d'Azur, BP 4229, 06304 Nice Cedex 4, France} \draft \date{\today}
\maketitle
\begin{abstract}
It is shown by perturbation techniques and numerical simulations that
the inverse cascade of kink-antikink annihilations, characteristic of
the Kolmogorov flow in the slightly supercritical Reynolds number
regime, is halted by the dispersive action of Rossby waves in the
$\beta$-plane approximation. For $\beta\to 0$, the largest excited
scale is $\propto \ln 1/\beta$ and differs strongly from what is
predicted by standard dimensional phenomenology which ignores
depletion of nonlinearity.
\end{abstract}
]
\def\pt{{\partial_t}}
\def\pdm{{\partial_x^{-4}}}
\def\pem{{\partial_x^{-5}}}
\def\pfm{{\partial_x^{-6}}}
\def\pcm{{\partial_x^{-3}}}
\def\pxxm{{\partial_x^{-2}}}
\def\pxm{{\partial_x^{-1}}}
\def\px{{\partial_x}}
\def\pxx{{\partial_x^2}}
\def\pc{{\partial_x^3}}
\def\pd{{\partial_x^4}}
\def\lc{{\lambda_3}}
\def\vfa{{\varphi_a}}
\def\vfb{{\varphi_b}}
\def\vpau{{\varphi_{a1}}}
\def\vpbu{{\varphi_{b1}}}
\def\vpad{{\varphi_{a2}}}
\def\vpbd{{\varphi_{b2}}}
\def\vvz{{\overline{v}^{(0)}}}
\def\vvu{{\overline{v}^{(1)}}}
\def\vvd{{\overline{v}^{(2)}}}
\def\Gr{{{\cal G}_{2}}}
\def\tv{{\tilde{v}}}
\def\Lrz{{{\cal L}_{0}}}
\def\Lru{{{\cal L}_{1}}}
\def\Lrd{{{\cal L}_{2}}}
Planetary-scale flow is subject to the competing effects of
quasi-two-dimensional turbulence and Rossby waves ($\beta$-effect).
It is known from phenomenological arguments and numerical simulations
that the inverse cascade which characterizes the large-scale dynamics
of two-dimensional turbulence in planetary flow can be halted by
Rossby wave dispersion and that the ensuing flow exhibits alternating
jets.  \cite{Rhin:75,Pane:93,Vall:93,Noza:97,Huan:98}. The standard
argument of Rhines \cite{Rhin:75} rests on a comparison between the
local eddy turnover time and the period of Rossby waves. More
generally, the interaction of waves and turbulence is a subject
with a wealth of applications in astro/geophysical flow and plasmas
(see, e.g., Ref.~\cite{zakharov}).

The case of strongly dispersive waves is reasonably well understood
through the theory of resonant wave interactions
(Refs.~\cite{zakharov,Fris:96} and references therein).  It is however
inapplicable to a rotating planet where the dispersive action of
Rossby waves (with a frequency $\beta/k$) is felt only at the very
largest scales (small wavenumber $k$). For the case of weakly
dispersive waves, the nonlinear dynamics in the absence of waves must
be understood in detail before we can find how they are affected by
the presence of the waves. We cannot resort just to dimensional
arguments which ignore the depletion of nonlinearities, an effect
which is very common in turbulence \cite{frisch} and which can take
extreme forms in some  situations, as we shall see. Our present study deals
with the Kolmogorov flow \cite{MS} for which the inverse cascade is
understood rather well in terms of kink-antikink dynamics
\cite{Kawa:82}. The basic Kolmogorov flow is ${\bf u} = (\cos y,0)$,
obtained by applying to the two-dimensional Navier--Stokes equation a
force ${\bf f}= \nu(\cos y,0)$.  The Kolmogorov flow has a large-scale
negative eddy viscosity instability when the kinematic viscosity $\nu$
is just below the critical value $\nu_c=1/\sqrt2$ \cite{MS}. The
large-scale dynamics are then, to leading order, one-dimensional
\cite{Fris:96,nepo}.  Close to $\nu_c$ and in the presence of Rossby
waves, the large-scale dynamics are governed by the one-dimensional
``$\beta$-Cahn--Hilliard'' equation with cubic non linearity, derived
by multiscale techniques in Ref.~\cite{Fris:96} (see also
Ref~\cite{nepo} for the case $\beta=0$), which reads
\begin{equation}
	\pt v = \lc \pxx U'(v) - \lambda_3 \pd v -\beta \pxm v.
\label{vbetach}
\end{equation}
Here, $U(v) = s^2(v^4/(2\Gamma^2)-v^2)$ is a quartic potential and
$\pxm$ denotes spatial integration for zero-average functions in the
interval $[0,L]$ over which periodicity is assumed. (In the original
setup of Ref.~\cite{Fris:96}, the constants  are
$s=1/\sqrt{3}$, $\Gamma=\sqrt{3/2}$ and $\lambda_3= 3/\sqrt2$.)

\begin{figure}
\iffigs
\centerline{\psfig{file=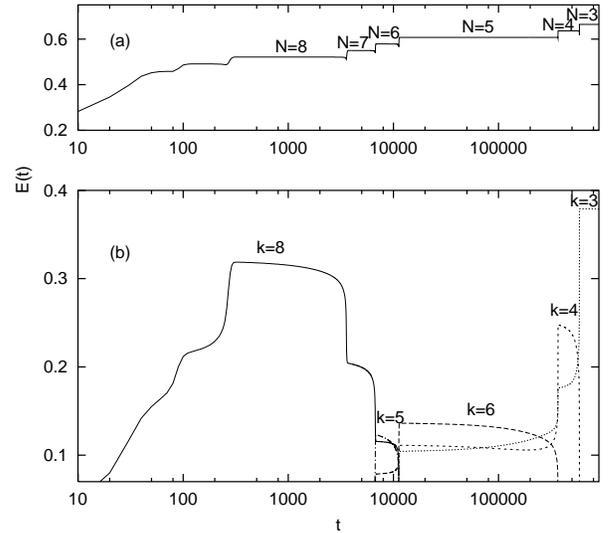,width=8cm,clip=}}
\fi
\caption{Simulation of the Cahn--Hilliard equation with $\beta =0$. 
(a)\,: evolution of the total energy. (b)\,: evolution of the the energies of
various Fourier modes (as labeled). An inverse cascade is observed
from $N=8$ to $N=3$. Eventually, the dominant mode becomes $N=1$ (not 
shown in the figure).} 
\label{CH}
\end{figure}
\begin{figure}
\iffigs
\centerline{\psfig{file=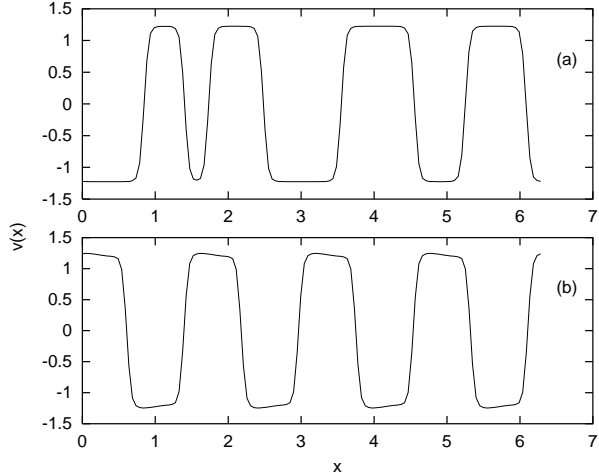,width=8cm,clip=}}
\fi
\caption{Solutions with four kink-antikink pairs. (a)\,: snapshot of
velocity $v(x,t)$ in a slowly evolving state for $\beta=0$; (b)\,:
final stable asymptotic state for $\beta=10^{-3}$ with $N=4$
exhibiting distorted plateaus between the kinks (this is actually a
traveling wave measured in a suitable moving frame).}
\label{sols}
\end{figure}

We begin with qualitative considerations illustrated by simulations.
For $\beta =0$, the solutions to this equation live essentially within
a slow manifold of soliton-like solutions with an alternation of
plateaus at the zeros $v=\pm \Gamma$ of the potential, separated by
alternating kinks and antikinks \cite{Bate:95}.  The corresponding
fixed points, having $N$ pairs of regularly spaced kinks and
antikinks, are all unstable saddle points of a Lyapunov functional,
except for $N=1$ which gives a stable absolute minimum.  The temporal
evolution is a cascade of annihilations of kink-antikink pairs,
leading eventually to the gravest $N=1$ mode \cite{Kawa:85}. This is
illustrated in Figs.~\ref{CH} and \ref{sols}(a) obtained by numerical
simulation using the method described in Ref.~\cite{Fris:96}. In the
Fourier space an inverse cascade is observed  with
the dominantly excited wavenumber shifting roughly to smaller and
smaller wavenumbers (see, e.g. Ref~\cite{She:87} and
Fig.~\ref{CH}(b)).  Except for the short kink-antikink annihilation
episodes, the motion of kinks is described by simple ODE's to be derived
below. They involve exponential couplings between adjacent kinks, so
that the typical duration of the plateaus increases
exponentially as $N$ decreases.
\vspace{-6mm}

\begin{figure}
\iffigs
\centerline{\psfig{file=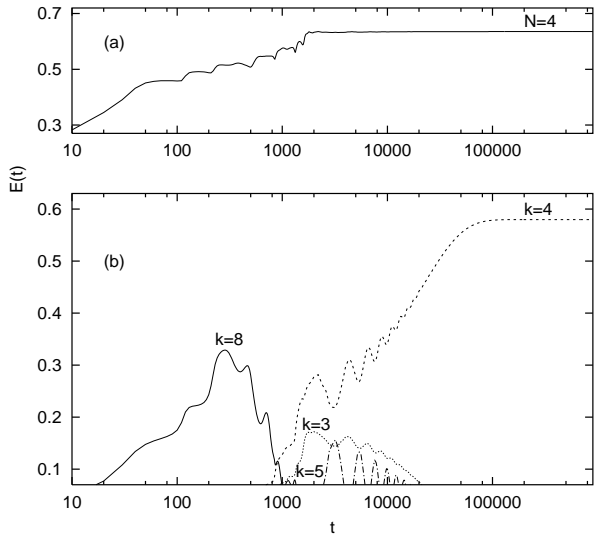,width=8cm,clip=}}
\fi
\caption{Same as Fig.~\ref{CH} but for $\beta= 10^{-3}$. Propagating 
Rossby waves accelerate the cascade and,  eventually, halt it when it
reaches  $N=4$ (four pairs of kink-antikinks).}
\label{cascB}
\end{figure}
For $\beta\ne 0$, the dispersive action of the waves modifies the
cascade. For small values of $\beta$, the solution retains the
characters of kink dynamics with superimposed propagating waves which
are expected to predominantly modify the dynamics at large scales
where the Rossby wave period is the smallest.  As the destabilizing
coupling between the kinks decreases exponentially with the typical
separation of adjacent kinks, the stabilizing effect of dispersive
waves can dominate for solutions of sufficiently large wavelength,
thereby halting the inverse cascade at an intermediate wavelength
$\Lambda=L/N$ with integer $N>1$.  This is illustrated with a
simulation for $\beta= 10^{-3}$ shown in Fig.~\ref{cascB}, where the
cascade is found to halt at wavenumber $k=4$, corresponding to a
wavelength $\Lambda=L/4$.

We turn now to more systematic theory, beginning again with
$\beta=0$, case for which we mostly follow Kawasaki and Ohta
\cite{Kawa:82}. The (pure) Cahn--Hilliard equation admits $M(x) = \pm
\Gamma \tanh sx$ as stationary solutions of $\pxx M_j - U'(M_j) = 0$.
When the typical separation between kinks is large, a solution with
$N$ kink-antikink pairs may be written, near the $j$th kink, as
$v(x,t)= M_j(x)+\tilde{v}_j(x,t)$, where $M_{j}(x) = \epsilon_{j}
M(x-x_{j})$ with $\epsilon_{j}=(-1)^j$ and exponentially small remainder
$\tilde{v}_j(x,t)$.  Hence, the time derivative
is, up to exponentially small terms, $\pt v(x,t) = -
\sum_{\ell=0}^{2N-1} \dot{x}_{\ell}(t) \px M_{\ell}(x) $. Substituting
into (\ref{vbetach}), integrating twice in space, multiplying by $\px
M_{j}$ and integrating over the periodicity interval $[0,L]$, we
obtain \FL
\begin{eqnarray}
\lefteqn{
\frac{1}{\lc} \sum_{\ell=0}^{2N-1} \dot{x}_{\ell} \int_0^L \int_0^L \px M_j(x)
\Gr(x-x') \px M_{\ell}(x') dx dx' = } \nonumber \\ && 
\int_0^L \frac{2 s^2}{\Gamma^2} \tilde{v}_{j}^2 (3 M_{j}  + 
\tilde{v}_{j}) \px M_j dx - 2 \epsilon_j \Gamma h(t) ,
\label{preode}
\end{eqnarray}
where $h(t)$ is a time-dependent  integration constant which can be
determined by the constraint of momentum conservation.
Carrying out the integrations in (\ref{preode}), we  obtain the 
ODE's for kink motion\,:
\begin{equation}
 {\cal A}_{j\ell} \dot{x_{\ell}} =  e^{-s(x_{j}-x_{j-1})} - 
 e^{-s(x_{j+1}-x_{j})}.
\label{kinkmotion}
\end{equation}
Here,  $x_{j}$ is the location of the $j$-th kink and ${\cal
A}_{j\ell}$ is a symmetric matrix given by
\[
{\cal A}_{j\ell} = \frac{1}{8 \lc s^{2}} \left( 
\epsilon_{j-\ell}\left(\Gr(x_j-x_{\ell})+ \frac{\pi^2}{12 L
s^2}\right) - \frac{1}{2 s}\delta_{j-\ell} \right),
\]
where $\Gr(x)$ is the $L$-periodic Green's function satisfying  $\pxx
\Gr(x) = - \delta(x)$. 
When $L \rightarrow \infty$,
$\Gr(x_j-x_{\ell})$ reduces to $-|x_j-x_{\ell}|/2$ and $\cal A$ is
easily inverted \cite{Kawa:85}.  Although the dynamics described by
(\ref{kinkmotion}) borrows most of its character from the heteroclinic
connection between unstable fixed points, the trajectory in
phase-space does not generally proceed from the vicinity of one fixed
point to another.  Fast jumps from one slow manifold to the next may
occur at distance from the fixed points. Thus, as the number of kinks
and antikinks decreases with time, the dominating wavenumber does not
necessarily decrease monotonically  as happens in the ``arithmetic''
inverse cascade of Ref.~\cite{She:87}. This may be seen in
Fig.~\ref{CH}(b) where the dominant mode goes from wavenumber
$k=8$, to 5 and then to 6.

The stability of a given fixed point $\vvz$ of (\ref{vbetach}) with 
respect to kink motion is 
obtained by differentiating (\ref{kinkmotion}). It is convenient to 
define the Fourier components $\psi_{m}$ of kink displacements 
$\delta x_{j}$ as
$\psi_{m} = (2N)^{-1} \sum_{m=0}^{2N-1} \delta x_{j} e^{-i \pi(mj/N)}$.
After some algebra, we obtain $\dot{\psi}_m = \sigma_{0} \psi_m$
for $m<N$, with the eigenvalue $\sigma_{0}$ given by 
\begin{equation}
  \sigma_{0} = \frac{128 s^{3} \lc e^{-s \Lambda}}{\Lambda} \sin^2 
  \theta_m \left( 1 - \frac{2(1 - \cos \theta_m)}{s \Lambda}\right)^{-1} ,
\label{s0}
\end{equation}
with $\theta_m = \pi m/N$ and  $\Lambda =L/N$
(the subscript $0$ refers to $\beta=0$).  
The corresponding eigenvectors for a given $m$ are $v_a(x) =
\sum_{j=0}^{2N-1} (-1)^j \cos j \theta_m \px M (x-x_{j})$ and $v_b(x)
= \sum_{j=0}^{2N-1} (-1)^j \sin j \theta_m \px M (x-x_{j})$.  The
leading order of (\ref{s0})  corrects a factor 2 error in
Ref.~\cite{Kawa:82} in which the exponential variation of
$\tilde{v}_j$ near $x_{j+1}$ and $x_{j-1}$ was neglected. Note that our result
is consistent  with  Ref.~\cite{Lang:71} and agrees with
our simulations. For large $\Lambda$, all the eigenvalues $\sigma_{0}$
(actually $\sigma_{0}(m)$) for $m>0$ are positive, thereby
demonstrating the instability of multi-kink-pair solutions.

We now turn to the case of non-vanishing small $\beta$, which
is studied by a singular perturbation method that we outline. The
stationary solution goes over into a (slowly) traveling wave solution
which, in a suitable frame moving with the velocity $c = \beta c_{1} +
\beta^2 c_{2} + O(\beta^{3})$, may be written in the time-independent
form $\overline{v} = \vvz + \beta \vvu + \beta^2 \vvd + O(\beta^3)$.
Here, $\vvu$ and $\vvd$ satisfy ${\cal F}\vvu =
Q^{(0)}$ and ${\cal F}\vvd = Q^{(1)}$ with ${\cal F}\equiv \pxx  - U_0''$, 
\begin{eqnarray}
&& Q^{(0)} \equiv \lc^{-1} \left(c_{1} \pxm
\vvz -  \pcm \vvz\right),\quad {\rm and}\,\,\,
Q^{(1)} \equiv\nonumber\\ &&\lc^{-1}\left (c_{1} \pxm \vvu
-  \pcm \vvu+ c_{2} \pxm \vvz\right) + \frac{1}{2} U_0'''
\left({\vvu}\right)^2,\nonumber
\end{eqnarray}
where $U_0\equiv U(\vvz)$.  At first order, $c_{1}$ is obtained from the solvability
condition $\int_{0}^{L} Q^{(0)} \px \vvz dx = 0$ as
\begin{equation}
  c_{1} = - \left(\frac{\Lambda^2}{48} - \frac{\pi^2}{12 s^2} +
  \frac{A}{\Lambda s^3}\right)\left(
  1 - \frac{4}{\Lambda s}\right)^{-1} ,
\label{c1}
\end{equation} 
where $A = 2.404113\cdots$.  Like $\sigma_{0}$ in (\ref{s0}),
$c_{1}$ is given in (\ref{c1}) up to exponentially small terms (for
large $\Lambda =L/N$) and has excellent agreement with numerical
simulations.  (Note that, at next order, $c_{2}$ vanishes.) The
first-order perturbation $\vvu$ of the traveling wave profile can also
be obtained perturbatively in a large-$\Lambda$ expansion (not given
here).  This leads to a distorsion of the plateaus, as seen in
Fig.~\ref{sols}(b).

The equation governing the perturbation $\delta v(x,t)\equiv
 v(x+ct,t)-\overline{v}(x)$ can be written  as
\begin{eqnarray}
&&\pt \varphi = {\cal L} \varphi,\qquad\varphi  \equiv \pxm \delta v,\nonumber\\
&&{\cal L}  \equiv - \lc \px (\pxx - U''(\overline{v})) \px + c \px - \beta
 \pxm.
\label{defLcal}
\end{eqnarray}
Unlike the case with $\beta = 0$, the perturbation to the stationary
solution of the $\beta$-CH equation does not reduce simply to a change
in the kink locations. We must also consider the dispersive effect of
the $\beta$-term which modifies the shape of the slow modes and
contributes to stability.  Substitution in (\ref{defLcal}) of the
expansions of $c$ and $\overline{v}$ in powers of $\beta$ leads to an
expansion ${\cal L} = \Lrz + \beta \Lru + \beta^2 \Lrd +
O(\beta^3)$. The unperturbed eigenvalue $\sigma_0$ becomes $\sigma =
\sigma_{0} + i \beta \mu_{1} + \beta^{2} \sigma_{2} + i \beta^{2}
\mu_{2} + O(\beta^{3})$.  Similarly, the eigenfunctions $\vfa = \pxm
v_a $ and $\vfb = \pxm v_b $ are respectively modified as $\vfa +
\beta \vpau + \beta^2 \vpad + O(\beta^3)$ and $\vfb + \beta \vpbu +
\beta^2 \vpbd O(\beta^3)$. A hierarchy of equations is then obtained
for $\vfa$, $\vfb$, $\vpau$,  $\vpbu$, $\vpad$, \ldots. As is usual in
such singular perturbation problems, the corrections to the
eigenvalues are obtained from solvability conditions. This gives
(with $\langle f, g \rangle \equiv L^{-1}\int_0^L f g dx$)
\begin{eqnarray}
&&\langle \vfb, \Lru \vfa \rangle = -\mu_{1}\langle
\vfb,\vfb \rangle,\nonumber\\
&& \langle \vfa, \Lru \vpau \rangle + \langle \vfa, \Lrd \vfa \rangle =
\sigma_{2} \langle \vfa, \vfa \rangle - \mu_{1} \langle \vfa,\vpbu
\rangle.\label{getsigma2}
\end{eqnarray}
The coefficient $\mu_1$ gives only the shift in the imaginary part of
the eigenvalue.  For the real part, the second-order term $\sigma_2$
is needed. It is obtained from (\ref{getsigma2}); limiting ourselves
to leading order large-$\Lambda$ contributions, we obtain after
considerable algebra
\begin{eqnarray}
\sigma_2 &=& - \frac{\Lambda^4}{69,\!120\; s^2 \lc}
\frac{q^2(4+9q^2)}{(1+q^2)^2} + O(\Lambda^{3}),
\label{s2r}\\
q&\equiv& \tan{ \pi m\over 2 N}.
\label{defq}
\end{eqnarray}
Note that the correction is negative and, hence, {\em stabilizing.}
Though the effect is small, it increases algebraically with $\Lambda$,
while the nonlinear coupling of kinks decreases exponentially in
(\ref{s0}).  Therefore, stabilization of the $m$-mode perturbation to
the stationary solution is obtained at leading order for large
$\Lambda$ when $\sigma_0+\sigma_2\beta^2<0$, that is
\begin{equation}
|\beta| > \beta_{c} 
= \sqrt{35,\!389,\!440\, \frac{e^{-s \Lambda}}{\Lambda^5} 
s^5 \lc^2 \frac{1 }{4 + 9 q^2} },
\label{betac}
\end{equation}
The condition is the most restrictive for $m=1$, that is $q= \tan 
\pi/(2N)$. 

When $\Lambda$ is only moderately large, an accurate estimate of
$\beta_{c}$ requires, in principle, a careful determination of $\vpbu$
and $\vpau$ by matching of solutions near to and away from the
kinks. We have used an alternative semi-numerical approach, which
works for all values of $\Lambda$, in which the auxiliary equations
stemming from the perturbation theory, are solved numerically with a
discretization over a quarter-wavelength of $\vfa$ and $\vfb$, using a
procedure written with {\em Mathematica}, available from the first
author (BL). The eigenvalues are obtained from the solvability
conditions of the discretized problem, where degeneracies are lifted
by imposing symmetries with appropriate boundary conditions for each
linear problem to be solved.  We have also solved directly the
stability problem by differentiating a Galerkin-truncated expansion of
(\ref{vbetach}) in Fourier modes.  Newton's algorithm is used to
obtain the traveling solution and its phase velocity; eigenmodes being
then calculated with a QR algorithm.  The latter procedure provides
both the fast and slow modes with exponential separation of the
eigenvalues. As the problem becomes very stiff when $L$ is large, we
are in practice limited to values of $L$ less than a few hundreds.

Table I compares the asymptotic and the semi-numerical solutions of the 
perturbation theory with the direct solution of the stability problem. 
There is excellent agreement between the direct calculation 
and the semi-numerical solution of the 
perturbation problem for $N < 5$. It breaks down when the 
approximation of separated kinks is no longer valid. The asymptotic
result (\ref{betac}) provides then the correct order of magnitude but is 
wrong by a factor 2 or 3. The convergence to the numerical solution
is observed at larger $L$ but is very slow\,: multiplying $L$ by 10 and 
50 narrows the discrepancy for mode 2 to $4.6\%$ and $1.3\%$,
respectively.

Numerical temporal integrations of (\ref{vbetach}) starting from 
random initial conditions eventually achieves a traveling wave
solution with one of the values of $N$ found stable by the
perturbation  theory. 


The main result of our perturbation theory is (\ref{betac}) which
gives the minimum value of $|\beta|$ able to stabilize a solution of
period $\Lambda =L/N$ with $N>1$. By inverting $\Lambda$ in terms of
$\beta$, we can explain the halting of the inverse cascade at a
wavelength $\Lambda$ which scales as $\Lambda= -(2/s)\ln |\beta|$ (to
leading order).
Note that standard phenomenology, based on dimensional
analysis \`a la Rhines \cite{Rhin:75} with equilibration of nonlinear
and Rossby characteristic times, gives a drastically different
scaling, namely $\Lambda \sim |\beta|^{-3}$. (The nonlinear time is
$\propto \Lambda ^2$ and the Rossby time to $1/(\beta\Lambda)$;
velocities are $O(1)$.) The discrepancy arises from the
failure of dimensional analysis to capture the almost complete
suppression of nonlinearity obtained in the plateaus.
\begin{table}
\begin{tabular}{l|ll|l|ll} 
& \multicolumn{2}{c}{direct} & \multicolumn{1}{c}{perturbation} & 
\multicolumn{2}{c}{perturbation} \\
&\multicolumn{2}{c}{calculation} & \multicolumn{1}{c}{\footnotesize(semi-numerical)} & 
\multicolumn{2}{c}{\footnotesize(large $\Lambda$-asymptotics)} \\
N & \multicolumn{1}{c}{$c_{1}$} &\multicolumn{1}{c} {$\beta_{c}$} & 
\multicolumn{1}{c}{$\beta_{c}$} & \multicolumn{1}{c}{$c_{1}$} & 
\multicolumn{1}{c}{$\beta_{c}$} \\ \hline
2 & $-35.002$ & $2.358 \,10^{-6}$ & $2.358 \,10^{-6}$ & $-35.002$  & $1.45 \,10^{-6}$ \\
3 & $-16.067$ & $4.965 \,10^{-4}$ & $4.966 \,10^{-4}$ & $-16.067$  & $2.20 \,10^{-4}$ \\ 
4 & $-9.2073$ & $1.069 \,10^{-2}$ & $1.067 \,10^{-2}$ & $-9.2090$  & $3.2 \,10^{-3}$ \\ 
5 & $-5.9507$ & $8.96  \,10^{-2}$ & $10.28 \,10^{-2}$ & $-5.9637$  & $1.8 \,10^{-2}$ \\ 
\end{tabular}
\vspace{2mm}
\caption{Comparison of the perturbation theory in both its asymptotic 
and numerical form with the direct stability calculation. The 
calculation is done for $L=76.93$ (10 unstable modes when $\beta=0$) 
and several values of $N$.}
\end{table}
The one-dimensional character of the large-scale dynamics of the
Kolmogorov flow in the slightly supercritical regime amplifies the
discrepancy. In multi-dimensional high-Reynolds numbers turbulence,
coherent structures (vortices, filaments, sheets, \ldots) exhibit also
strongly depleted nonlinearities which dimensional arguments fail to
capture. It would be of interest to study dispersive stabilization for
a strongly nonlinear high-Reynolds number inverse cascade of the kind
considered by Kraichnan \cite{kraichnan}. So far no systematic theory
can handle this, although some progress has been made recently on a
related passive scalar problem \cite{GV}.

Finally, we mention that in recent work, devoted to zonal jets in
planetary flow, Manfroi and Young~\cite{Manf:98} studied a closely
related problem in which stabilization is provided by a friction term
$-r v$ added to the right hand side of (\ref{vbetach}). We find that a
suitable adaptation of the analysis for the dispersive case gives
stabilization for $r >r_c \sim e^{-s\Lambda}/\Lambda$.

We are grateful to Joanne Deval for careful checking of all the
calculations and to an anonymous referee for useful remarks.

\end{document}